\documentclass[article]{revtex4}
\usepackage{amsfonts}
\usepackage{amssymb}
\usepackage{graphicx}

\begin{document}

\def\be{\begin{equation}}
\def\ee{\end{equation}}
\def\bea{\begin{eqnarray}}
\def\eea{\end{eqnarray}}
\def\tr{{\rm tr}\, }
\def\nn{\nonumber \\}
\def\e{{\rm e}}

\title{Oscillating Universe from inhomogeneous EoS and coupled dark energy}

\author{Diego S\'{a}ez-G\'{o}mez$^{1,}$\footnote{Electronic address:saez@ieec.uab.es}}
\affiliation{$^{1}$Consejo Superior de Investigaciones Cient\'\i ficas
ICE/CSIC-IEEC, Campus UAB, Facultat de Ci\`encies, Torre
C5-Parell-2a pl, E-08193 Bellaterra (Barcelona) Spain}
\begin{abstract}
An occurrence of an oscillating Universe is showed using an inhomogeneous equation of state for dark energy fluid. The Hubble parameter described presents a periodic behavior such that  early and late time acceleration are unified under the same mechanism. Also, it is considered a coupling between dark energy fluid, with homogeneous and constant EoS, and matter, that gives a periodic Universe too. The possible phantom phases and future singularities are studied in the oscillating Universe under discussion. The equivalent scalar-tensor representation for the same oscillating Universe is presented too.  
\end{abstract}
\maketitle

\section{INTRODUCTION}
The discovery of cosmic acceleration in 1998 by two groups independently\cite{DiscAcc}-\cite{DiscAcc2} brought to propose a big number of dark energy models (for recent reviews, see Ref. \cite{DErev}-\cite{DErev2}), where this mysterius cosmic fluid was introduced under the prescription that its equation of state (EoS) parameter should be less than -1/3. At the era of precision cosmology, the observational data establishes that the EoS parameter for dark energy $w$ is close to -1 (see Ref.\cite{obsData}-\cite{obsData1}). The main task is to describe the nature of this component, for that purpose, several candidates have been proposed, we can mention  the cosmological constant model with $w=-1$, the so-called dark fluids with an inhomogeneous EoS\cite{InhEoS}-\cite{InhEoSandOscillating1},  and the  quintessence/phantom scalar fields models\cite{phantom}-\cite{scalarth3}. This kind of models may reproduce late-time acceleration but it is not easy to construct a model that keeps untouched the radiation/matter dominated epochs.\\  
An additional gain of these models with scalar and ideal dark fluids is  that they allow the possibility to unify early and late time acceleration under the same mechanism, in such a way that the Universe history may be reconstructed completly. On the other hand, it is impotant to keep in mind that these models represent just an effective description that own a number of well-known problems, as the ending of inflation. Nevertheless, they may represent a simple and natural way to resolve the coincidence problem, one of the possibilities may be an oscillating Universe (Ref. \cite{InhEoSandOscillating} and \cite{OscillaUniverse}-\cite{oscillate4}), where the differents phases of the Universe are reproduced due to its periodic behavior. The purpose of this paper is to show that, from inhomogeneous EoS for a dark energy fluid, an oscillating Universe is obtained, and several examples are given to illustrate it. It is studied the possibility of an interaction between dark energy fluid, with homogeneous EoS, and matter that also reproduces that kind of periodic Hubble parameter, such case has been studied and is allowed by the observations (see \cite{CopuplingObsv}). The possible phantom epochs are explored, and the possibility that Universe may reach a Big Rip singularity (for a classification of future singularities, see Ref. \cite{CouplingAndSingula}).  \\
The organization of this paper is the following: in Sec. 2, a dark energy fluid with inhomogeneous EoS is presented, where this EoS depends on Hubble parameter, its derivatives and on time, it is showed that this kind of EoS reproduces an oscillating behavior of the Hubble parameter. In Sec. 3 a matter component is included,  in the first part the problem is driven supossing no coupling between the matter and the dark fluid, a periodic Hubble parameter is obtained under some restriction on inhomogeneous EoS for the dark fluid, and in the second part a coupling is introduced, and it is showed that for a constant homogeneous EoS for dark energy, it is possible reconstruct early and late-time acceleration in a naturally way, due to the interaction between both fluids. Finnally, in Sec. 4 the mathematically equivalent scalar tensor description is showed, where the above solutions are reproduced by canonical/phantom scalar fields.  

\section{INHOMOGENEOUS EQUATION OF STATE FOR DARK ENERGY}

\begin{figure}
 \centering
 \includegraphics[width=3in,height=2in]{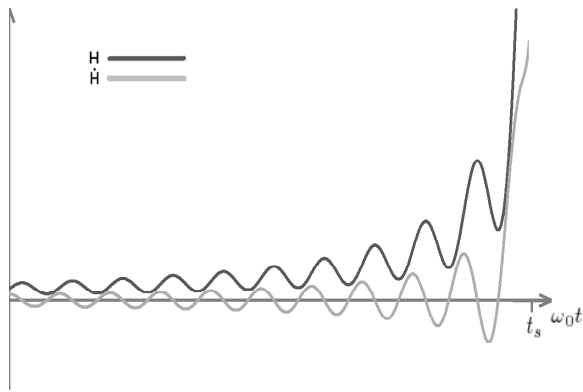}
 
 \caption{The Hubble parameter H and $\dot{H}$ for a value $w=-1.1$. Phantom phasees ocurrs periodically, and a Big Rip singularity takes place at Rip time $t_s$.}
 \label{fig1}
\end{figure}

Let us consider firstly a Universe filled with a dark energy fluid, neglecting the rest of possible components (dust matter, radiation..), where its EoS depends on the Hubble parameter and its derivatives, such kind of EoS has been treated in several articles\cite{InhEoS}-\cite{InhEoSandOscillating1}. We show that for some choices of the EoS, an oscillating Universe resulted\cite{OscillaUniverse}-\cite{oscillate4}, which may include phantom phases. Then, the whole Universe history, from inflation to cosmic acceleration, is reproduced in such a way that observational constraints may be satisfied\cite{obsData}-\cite{obsData1}. We work in a spatially flat FRW Universe , then the metric is given by:
\be 
ds^2=-dt^2+a(t)^2\sum_{i=1}^{3}dx_i^2\ .
\label{1.1}
\ee
The Friedmann equations are obtained:
\be
H^2=-\frac{\kappa^2}{3}\rho, \quad \quad \dot{H}=-\frac{\kappa^2}{2}\left( \rho+p\right)\ . 
\label{1.2}
\ee
By combining the FRW equations, the energy conservation equation for dark energy density results:
\be
\dot{\rho}+3H(\rho+ p)=0\ .
\label{1.3}
\ee
At this section, the EoS considered is given to have the general form:
\be
p= w\rho+g(H, \dot{H}, \ddot{H},..;t)\ ,
\label{1.4}
\ee
where $w$ is a constant and $g(H, \dot{H}, \ddot{H},..;t)$ is an arbitrary function of the Hubble parameter $H$, its derivatives and the time $t$, (such kind of EoS has been treated in Ref. \cite{InhEoS}). Using the FRW equations (\ref{1.2}) and (\ref{1.4}), the following  differential equation is obtained:
\be
\dot{H}+\frac{3}{2}(1+w)H^2+\frac{\kappa^2}{2}g(H, \dot{H}, \ddot{H},..;t)=0\ .
\label{1.5}
\ee
Hence, for a given function $g$, the Hubble parameter is determinated by solving the equation (\ref{1.5}). It is possible to reproduce an oscillating Universe by an specific EoS (\ref{1.4}). To illustrate this construction, let us consider the following $g$ function as an example:
\be
g(H, \dot{H}, \ddot{H})=-\frac{2}{\kappa^2}\left( \ddot{H}+\dot{H}+\omega_0^2H +\frac{3}{2}(1+w))H^2-H_0\right)\ , 
\label{1.6}
\ee
 where $H_0$ and $\omega_0^2$ are constants. By substituting (\ref{1.6}) in (\ref{1.5}) the Hubble parameter equation acdquieres the form:
\be
\ddot{H}+\omega_0H=H_0\ ,
\label{1.7}
\ee
which is the classical equation for an harmonic oscillator. The solution\cite{scalar2} is found:
\be
H(t)=\frac{H_0}{\omega_0^2}+H_1 \sin(\omega_0t+\delta_0)\ ,
\label{1.8}
\ee
where $H_1$ and $\delta_0$ are  integration constants. To study the system, we calculate the first derivative of the Hubble parameter, which is given by $\dot{H}=H_1\cos(\omega_0t+\delta_0)$, so the Universe governed by the dark energy fluid (\ref{1.6}) oscillates between phantom and non-phantom phases with a frecuency given by the constant $\omega_0$, constructing inflation epoch and  late-time acceleration under the same mechanism, and Big Rip singularity avoided \\
As another example, we consider the following EoS (\ref{1.4}) for the dark energy fluid:
\be
p= w\rho+\frac{2}{\kappa^2}Hf'(t)\ .
\label{1.9}
\ee
In this case  $g(H;t)=\frac{2}{\kappa^2}Hf'(t)$, where $f(t)$ is an arbitrary function of the time $t$, and the prime denotes a derivative on $t$. The equation (\ref{1.5}) takes the form:
\be
\dot{H}+Hf'(t)=-\frac{3}{2}(1+w)H^2\ .
\label{1.10}
\ee
This is the well-known Bernoulli differential equation. For a function $f(t)=-ln\left(H_1 + H_0 \sin\omega_0t \right)$, where $H_1>H_0$ are arbitrary constants, then the following solution for (\ref{1.10}) is found:
\be
H(t)=\frac{H_1+H_0 \sin \omega_0t}{\frac{3}{2}(1+w)t+k}\ ,
\label{1.11}
\ee
here, the $k$ is an integration constant. As it is seen, for some values of the free constant parameters, the Hubble parameter tends to infinity for a given finite value of $t$.  The first derivative of the Hubble parameter is given by: 
\be
\dot{H}=\frac{\frac{H_0}{\omega_0}(\frac{3}{2}(1+w)t+k)\cos\omega_0t-(H_1+H_0 \sin \omega_0t)\frac{3}{2}(1+w)}{(\frac{3}{2}(1+w)t+k)^2}\ .
\label{1.12}
\ee
As it is shown in fig.\ref{fig1}, the Universe has a periodic behavior, it passes through phantom and non-phantom epochs, with its respectives transitions. A Big Rip singularity may take place depending on the value of $w$, such that it is avoided for $w\geq-1$, while if $w<-1$ the Universe reaches the singularity in the  Rip time given by $t_s=\frac{2k}{3|1+w|}$.

\section{ DARK ENERGY IDEAL FLUID AND DUST MATTER}
\subsection{No coupling between matter and dark energy}
\begin{figure}
 \centering
 \includegraphics[width=3in,height=2in]{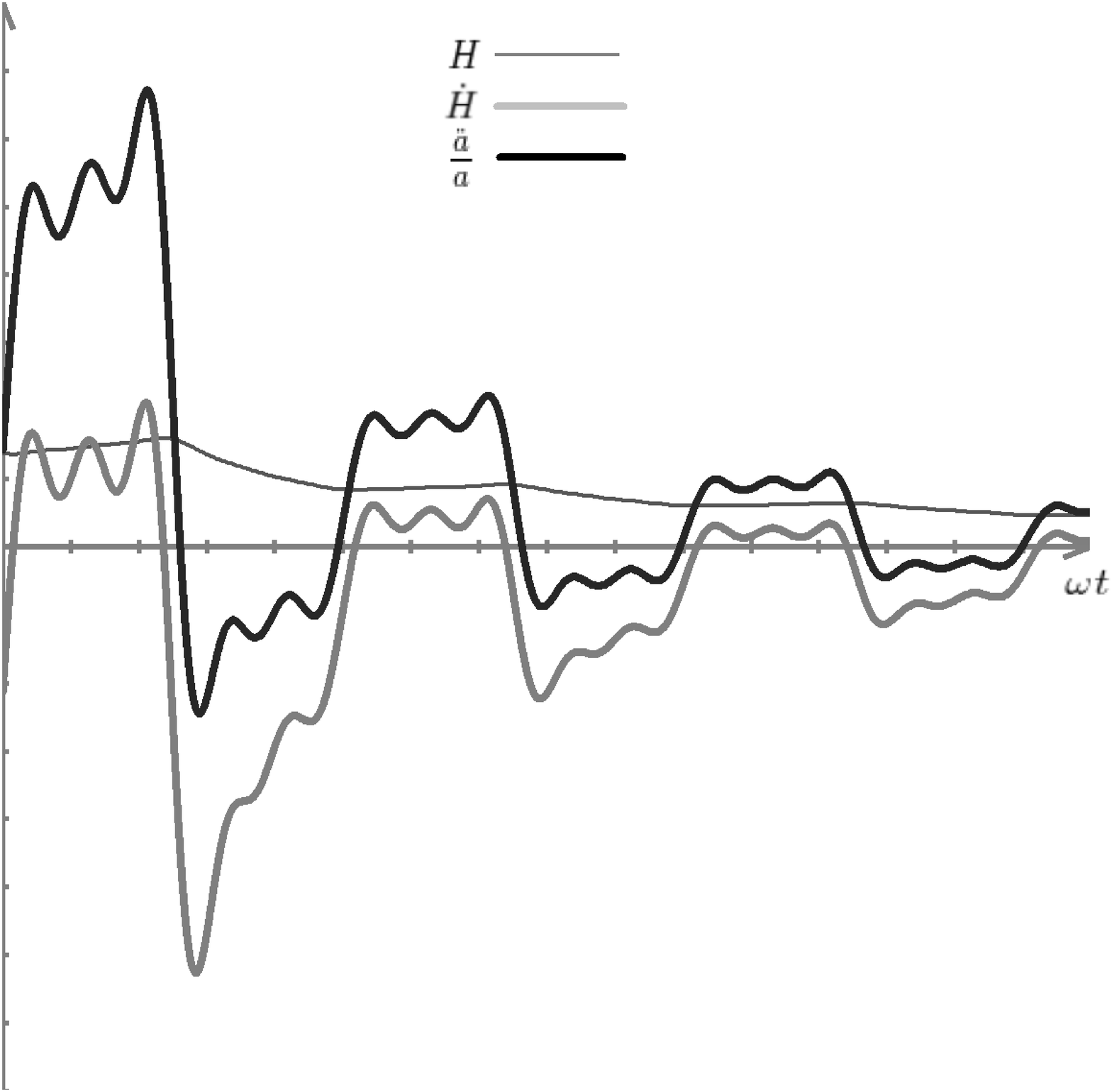}
 
 \caption{The Hubble parameter, its derivative and the acceleration parameter are represented for the ``step''  model for values $w_0=-0.2$ and $w_1=1$.}
 \label{fig:2}
\end{figure}

Let us now explore a more realistic model by introducing a matter component with EoS  given by $p_m=w_m\rho_m$, we consider an inhomogeneous EoS for the dark energy component\cite{InhEoS}$^-$\cite{InhoEoSandCoupling}. It is shown below that an oscillating Universe  may be obtained by constructing an specific EoS. In this case, the FRW equations (\ref{1.2}) take the form:
\be
H^2=-\frac{\kappa^2}{3}(\rho+\rho_m), \quad \quad \dot{H}=-\frac{\kappa^2}{2}\left( \rho+p+\rho_m+p_m\right)\ . 
\label{1.13}
\ee
At this section, we consider a matter fluid that doesnt interact with the dark energy fluid, then the energy conservation equations are satisfied for each fluid separately:  
\be
\dot{\rho_m}+3H(\rho_m+ p_m)=0, \quad \quad \dot{\rho}+3H(\rho+ p)=0\ .
\label{1.14}
\ee
It is useful to construct an specific solution for the Hubble parameter by defining the effective EoS with an effective parameter $w_{eff}$:
\be
w_{eff}=\frac{p_{eff}}{\rho_{eff}}, \quad \rho_{eff}=\rho+\rho_m, \quad p_{eff}=p+p_m\ ,
\label{1.15}
\ee
and the energy conservation equation $\dot{\rho}_{eff}+3H(\rho_{eff}+ p_{eff})=0 $ is satisfied. We consider a dark energy fluid which is described by the EoS describes by the following expression:
\be
p=-\rho+\frac{2}{\kappa^2}\frac{2(1+w(t))}{3\int(1+w(t))dt}-(1+w_m)\rho_{m0}\e^{-3(1+w_m)\int dt \frac{2}{3\int (1+w(t))}}\ ,
\label{1.16}
\ee
here $\rho_{m0}$ is a constant, and $w(t)$ is an arbitrary function of time $t$. Then the following solution is found:
\be
H(t)=\frac{2}{3\int dt(1+w(t))}\ . 
\label{1.17}
\ee
And the effective parameter (\ref{1.15}) takes the form $w_{eff}=w(t)$. Then, it is shown that a solution for the Hubble parameter may be constructed from EoS (\ref{1.16}) by specifying a function $w(t)$. \\
\\Let us consider an example\cite{OscillaUniverse} with the following function for $w(t)$:
\be
w=-1+w_0\cos \omega t\ .
\label{1.18}
\ee
In this case, the EoS for the dark energy fluid, given by (\ref{1.16}), takes the form:
\be
p=-\rho +\frac{4}{3\kappa^2}\frac{\omega w_0\cos\omega t}{w_1+w_0\sin\omega t}-(1+w_m)\rho_{m0}\e^{-3(1+w_m)\frac{2w}{3(w_1+w_0 \sin \omega t)}}\ ,
\label{1.19}
\ee
where $w_1$ is an integration constant. Then, by (\ref{1.17}), the Hubble parameter yields:
\be
H(t)=\frac{2\omega}{3(w_1+w_0 \sin \omega t)}\ .
\label{1.20}
\ee 
The Universe passes through phantom and non phantom phases since the first derivative of the Hubble parameter has the form:
\be
\dot{H}=-\frac{2\omega^2 w_0\cos\omega t}{3(w_1+w_0\sin\omega t)^2}\ .
\label{1.21}
\ee
In this way, a Big Rip singularity will take place in order that $\vert w_1\vert<w_0$, and it is avoided when $\vert w_1\vert>w_0$. As it is shown, this model reproduces unified inflation and cosmic acceleration in a natural way, where the Universe presents a periodic behavior. In order to reproduce accelerated and decelerated phases, the acceleration parameter is studied, which is given by:
\be
\frac{\ddot{a}}{a}=\frac{2\omega^2}{3(w_1+w_0 \sin \omega t)^2}\left( \frac{2}{3}-w_0\cos\omega t\right) \ .
\label{1.21bis}
\ee
Hence, if $w_0>2/3$ the differents phases that Universe passes are reproduced by the EoS (\ref{1.19}), presenting a periodic evolution that may unify all the epochs by the same description. 
\\
\\As a second example, we may consider a classical periodic function, the step function:
\begin{equation}
w(t)=-1+
\left\{ 
\begin{array}{lr}
w_{0}&0<t<T/2\\
w_{1}&T/2<t<T 
\end{array}
\right.\ ,
\label{1.22}
\end{equation}
and $w(t+T)=w(t)$. It is useful to use a Fourier expansion such that the function (\ref{1.22}) become continuos. Aproximating to third order, $w(t)$ is given by:
\be
w(t)=-1+\frac{(w_0+w_1)}{2}+\frac{2(w_0-w_1)}{\pi}\left( \sin\omega t +\frac{\sin3\omega t}{3}+\frac{\sin5\omega t}{5}\right) \ .
\label{1.23}
\ee
Hence, the EoS for the Dark energy ideal fluid is given by (\ref{1.16}), and the solution (\ref{1.17}) takes the following form:
\[
H(t)=\frac{2}{3} \left[w_2+\frac{(w_0+w_1)}{2}t \right.   
\]
\be
\left. -\frac{2(w_0-w_1)}{\pi\omega}\left( \cos\omega t+ \frac{\cos3\omega t}{9}+\frac{\cos5\omega t}{25}\right) \right]^{-1}\ .
\label{1.24}
\ee
The model is studied by the first derivative of the Hubble parameter in order to see the possible phantom epochs, since:
\[
\dot{H}=-\frac{3}{2}H^2\left[ \frac{(w_0+w_1)}{2} \right.
\]
\be
\left. -\frac{2}{\pi}(w_0-w_1)\left( \sin\omega t+\frac{\sin3\omega t}{3}+\frac{\sin5\omega t}{5}\right)  \right] \ .
\label{1.25}
\ee 
Then, depending on the values from $w_0$ and $w_1$ the Universe passes through phantom phases. To explore the different epochs of acceleration and deceleration that the Universe passes on,  the acceleration parameter is calculated:
\[
\frac{\ddot{a}}{a}=H^{2}+\dot{H}=
\]
\be
H^2\left[1-\frac{3}{2}\left( \frac{(w_0+w_1)}{2}-\frac{2}{\pi}(w_0-w_1)\left( \sin\omega t+\frac{\sin3\omega t}{3}+\frac{\sin5\omega t}{5}\right)\right)  \right]\ .   
\label{1.26}
\ee
Then, in order to get  acceleration and deceleration epochs, the constants parameters $w_0$ and $w_1$ may be chosen such that $w_0<2/3$ and $w_1>2/3$, as it is seen by (\ref{1.22}). For this selection, phantom epochs take place in the case that $w_0<0$. In any case, the oscillated behavior is damped by the inverse term on the time $t$, as it is shown in fig.\ref{fig:2}, where the acceleration parameter is ploted for some determinated values of the free parameters. This inverse time term makes reduce the acceleration and the Hubble parameter such that the model tends to a static Universe. \\

We  consider now a third example where a classical damped oscillator is showed, the function $w(t)$  is given by:
\be
w(t)=-1+\e^{-\alpha t}w_0\cos\omega t\ ,
\label{1.27}
\ee
here $\alpha$ and $w_0$ are two positive constants. Then, the EoS for the dark energy ideal fluid is constructed from (\ref{1.16}). The solution for the Hubble parameter (\ref{1.17}) is integrated, and takes the form:
\be
H(t)=\frac{2}{3}\frac{\omega^2+\alpha^2}{w_1+w_0\e^{-\alpha t}(\omega\sin\omega t-\alpha\cos\omega t)}\ ,
\label{1.28}
\ee
where $w_1$ is an integration constant. The Hubble parameter oscillates damped by an exponential term, and for big times, it tends to a constant $H(t\longrightarrow\infty)=\frac{2}{3}\frac{\omega^2+\alpha^2}{w_1}$, recovering the cosmological constant model. The Universe passes through different phases as it may be shown by the accelerated parameter:
\be
\frac{\ddot a}{a}=H^2\left(1-\frac{3}{2}\e^{-\alpha t}w_0\cos\omega t\right)\ .
\label{1.29}
\ee
It is possible to restrict $w_0>\frac{2}{3}$ in order to get deceleration epochs when the matter component dominates. On the other hand, the Universe also passes through phantom epochs, since the Hubble parameter derivative gives:
\be
\dot{H}=-\frac{3}{2}H^2\e^{-\alpha t}w_0\cos\omega t\ .
\label{1.30}
\ee
Hence, the example (\ref{1.27}) exposes an oscillating Universe with a frecuency given by $\omega$ and damped by a negative exponential term, which depends on the free parameter $\alpha$, these may be adjusted such that the phases agree with the phases times constraints by the observational data. 

\subsection{Dark energy and coupled matter}
In general, one may consider a Universe filled with a dark energy ideal fluid whose Eos is given $p=w\rho$, where $w$ is a constant, and matter described by $p_m=\omega_m\rho_m$,  both interacting with eah other. In order to preserve the energy conservation, the equations for the energy density are written as following:
\be
\dot{\rho_m}+3H(\rho_m+p_m)= Q, \quad \quad
\dot{\rho}+3H(\rho+p_)= -Q\ ,
\label{1.31}
\ee
here $Q$ is an arbitrary function. In this way, the total energy conservation is satisfied $\dot{\rho}_{eff}+3H(\rho_{eff}+p_{eff})=0$, where $\rho_{eff}=\rho+\rho_m$ and $p_{eff}=p+p_m$, and the FRW equations (\ref{1.13}) doesnt change. To resolve this set of equations for a determined function $Q$, the second FRW equation (\ref{1.13}) is combined with the conservation equations (\ref{1.31}), this yields:
\[
\dot{H}=-\frac{\kappa^2}{2}\left[(1+w_m)\frac{\int Q\exp(\int dt3H(1+w_m))}{\exp(\int dt3H(1+w_m))} \right.
\]
\be
\left.+(1+w)\frac{-\int Q\exp(\int dt3H(1+w))}{\exp(\int dt3H(1+w))} \right]\ .
\label{1.32}
\ee 

In general, this is difficult to resolve for a function Q. As a particular simple case is the cosmological constant where the dark energy EoS parameter $w=-1$ is considered,  the equations become very clear, and  (\ref{1.31}) yields $\dot{\rho}=-Q$, which is resolved and the  dark energy density is given by:
\be
\rho(t)=\rho_0-\int dt Q(t)\ ,  
\label{1.33} 
\ee
where $\rho_0$ is an integration constant. Then, the Hubble parameter is obtained by introducing (\ref{1.33}) in the FRW equations, which yields:
\be
\dot{H}+\frac{3}{2}(1+w_m)H^2=\frac{\kappa^2}{2}(1+w_m)\left(\rho_0-\int dt Q \right)\ . 
\label{1.34}
\ee
Hence, Hubble parameter depends essentially on the form of the coupling function $Q$. This means that a Universe model may be constructed from the coupling between matter and dark energy fluid, which is given by $Q$, an arbitrary function. It is showed below that some of the models given in the previus section by an inhomogeneous EoS dark energy fluid, are reproduced by a dark energy fluid with constant EoS ($w=-1$), but coupled to dust matter. By differentiating equation (\ref{1.34}), the function $Q$ may be written in terms of the Hubble parameter and its derivatives: 
\be
Q=-\frac{2}{\kappa^2}\frac{1}{1+w_m}\left(\ddot{H}+3(1+w_m)H\dot{H} \right)\ .
\label{1.35}
\ee 
\\
As an example, we use the solution(\ref{1.8}):
\be
H(t)=H_0+H_1 \sin(\omega_0t+\delta_0)\ .
\label{1.36}
\ee
Then, by the equation (\ref{1.34}), the function $Q$ is given by:
\be
Q(t)=\frac{2}{\kappa^2(1+w_m)}\left[ H_0\omega^2\sin\omega t +3(1+w_m)h_0\omega\cos\omega t(H_1+H_0\sin\omega t)) \right] \ . 
 \label{1.37}
\ee
Then, the oscillated model (\ref{1.36}) is reproduced by a coupling between matter and dark energy, which also oscillates. Some more complicated models may be constructed for complex functions $Q$. As an example let us consider the solution (\ref{1.20}):
\be
H(t)=\frac{2\omega}{3(w_1+w_0 \sin \omega t)}\ .
\label{1.38}
\ee 
The coupling function (\ref{1.35}) takes the form:
\[
Q(t)=-\frac{4}{3\kappa^2}\frac{\omega^3 w_0}{(1+w_m)(w_1+w_0 \sin \omega t)^3}
\]
\be
\left[\sin\omega t(w_1+w_0 \sin \omega t)^2+2w_0\cos^2\omega t-2(1+w_m)\cos\omega t \right]\ .
\label{1.39}
\ee
This coupling function reproduces a oscillated behavior that unifies the different epochs in the Universe. Hence, it have been showed that for a constant EoS for the dark energy with $w=-1$, inflation and late-time acceleretion are given in a simple and natural way. 

\section{SCALAR-TENSOR DESCRIPTION}
Let us now consider the solutions showed in the last sections through  scalar-tensor description, such equivalence has been constructed in Ref. \cite{scalar-th-fluids}. We assume, as before, a flat FRW metric, a Universe filled with a ideal matter fluid with EoS given by $p_m=w_m\rho_m$, and no coupling between matter and the scalar field. Then, the following action is considered:
\be
S=\int dx^{4}\sqrt{-g}\left[ \frac{1}{2\kappa^{2}}R
 - \frac{1}{2} \omega (\phi)
\partial_{\mu} \phi \partial^{\mu }\phi -V(\phi )+L_{m}\right]\ ,
\label{1.40}
\ee
here $\omega(\phi)$ is the kinetic term and $V(\phi)$ represents the scalar potential. Then, the corresponding FRW equations are written as:
\be
H^{2} = \frac{\kappa ^{2}}{3}\left( \rho _{m}
+\rho_{\phi}\right)\ , \quad \quad \dot H = -\frac{\kappa ^{2}}{2}\left(
\rho _{m}+p_{m}+\rho _{\phi }+p_{\phi }\right)\ ,
\label{eq:1.41}
\ee 
where $\rho_{\phi}$ and $p_{\phi}$ given by:
\be
\rho _{\phi } = \frac{1}{2} \omega (\phi )\, {\dot \phi}^{2}
+V(\phi)\ ,\quad \quad
p_{\phi } = \frac{1}{2} \omega (\phi ) \, {\dot \phi}^{2}-V(\phi)\ .
\label{eq:1.42}
\ee
By assuming:
\bea
\omega (\phi ) = -\frac{2}{\kappa ^{2}}f^{\prime }(\phi )
 -(w_{m}+1)F_{0} \e^{-3(1+w_{m})F(\phi )}\ , \nn 
&&\nonumber \\
V(\phi ) = \frac{1}{\kappa ^{2}}
\left[ {3f(\phi)}^{2}+f^{\prime }(\phi ) \right]
+\frac{w_{m}-1}{2}F_{0}\, \e^{-3(1+w_{m})F(\phi )}\ .
\label{eq:1.43}
\eea
The following solution is found\cite{scalar2}$^-$\cite{scalarth3}:
\be
\phi =t\ , \quad H(t)=f(t)\ ,
\label{1.44}
\ee
which yields:
\be
a(t)=a_{0}\e^{F(t)}, \qquad a_{0}=\left(
\frac{\rho _{m0}}{F_{0}}\right) ^{\frac{1}{3(1+w_{m})}}.
\label{eq:1.45}
\ee
Then, we may assume solution (\ref{1.28}), in such case the $f(\phi)$ function takes the form:
\be
f(\phi)=\frac{2}{3}\frac{\omega^2+\alpha^2}{w_1+w_0\e^{-\alpha \phi}(\omega\sin\omega \phi-\alpha\cos\omega \phi)}\ ,
\label{1.46}
\ee
And by (\ref{eq:1.43}) the kinetic term and the scalar potential are given by:
\bea
\omega(\phi)=\frac{3}{\kappa^2}f^2(\phi)w_0\e^{-\alpha\phi}\cos\omega\phi-(1+w_m)F_0\e^{-3(1+w_m)F(\phi)}, \nn
V(\phi)=\frac{3f^2(\phi)}{\kappa^2}\left(1-\frac{1}{2}w_0\e^{-\alpha\phi}\cos\omega\phi \right) +\frac{w_m-1}{2}F_0\e^{-3(1+w_m)F(\phi)},
\label{1.47}
\eea
where $F(\phi)=\int d\phi f(\phi)$ and $F_0$ is an integration constant. Then, the periodic solution (\ref{1.28}) is reproduced in the mathematical equivalent formulation in scalar-tensor theories by the action (\ref{1.40}) and  explicit kinetic term and scalar potential, in this case, is given by (\ref{1.47}).
\section{DISCUSSIONS}
Along the paper, it has been presented a Universe model that reproduces in a natural way the early and late-time acceleration by a periodic behavior of the Hubble parameter.  The  late-time transitions are described by this model: the transition from deceleration to acceleration, and the possible transition from non-phantom to phantom epoch. The observational data does not restrict yet the nature and details of the EoS for dark energy, then the possibility that Universe behaves periodically is allowed. For that purpose, several examples have been studied in the present paper, some of them driven by an inhomogeneous EoS for dark energy, and others by a coupling between dark energy and matter which also may provide another possible constraint to look for. On the other hand, as it was commented at the introduction, one has to keep in mind that these kind of models, scalar theories or dark fluids, should be checked to probe if the unification presented is realistic, especially in order to reproduce the details of inflation, to realize the perturbations structure... However, that task is beyond the scopes of the present paper, whose main objective is to show the possibility of the reconstruction of an oscillating Universe from the descriptions detailed.

\begin{acknowledgments}

 I thank  Emilio Elizalde and Sergei Odintsov for suggesting this problem, and for giving the ideas and fundamental information to carry out this task. This work was supported  by MEC (Spain), project FIS2006-02842, and in part by project PIE2007-50/023.

\end{acknowledgments}

\end{document}